\documentclass[doublespacing]{elsart}

  \usepackage{graphicx}

\usepackage{amssymb}

\begin{document}

\begin{frontmatter}



\title{Analytical Potential Energy Function for the Ground State $X^{1}\Sigma^{+}$ of LaCl}

\author{Lin-Hong Chen\corauthref{cor1}},
\corauth[cor1]{Corresponding author. Supported by NSFC, grant no. 10074037.}
\ead{chenlh98g@mails.tsinghua.edu.cn}
\author{Ren-Cheng Shang}
\address{Center for Astrophysics, Tsinghua University, Beijing 100084, China}

\begin{abstract}
The equilibrium geometry, harmonic frequency and dissociation energy of lanth\-anum monochloride have been calculated at B3LYP, MP2, QCISD(T) levels with energy-consistent relativistic effective core potentials. The possible electronic state and reasonable dissociation limit for the ground state are determined based on atomic and molecular reaction statics. Potential energy curve scans for the ground state $X^{1}\Sigma^{+}$ have been carried out with B3LYP and QCISD(T) methods due to their better performance in bond energy calculations. We find the potential energy calculated with QCISD(T) method is about 0.5 eV larger than dissociation energy when the diatomic distance is as large as 0.8 nm. The problem that single-reference ab initio methods don't meet dissociation limit during calculations of lanthanide heavy-metal elements is analyzed. We propose the calculation scheme to derive analytical Murrell-Sorbie potential energy function and Dunham expansion at equilibrium position. Spectroscopic constants got by standard Dunham treatment are in good agreement with results of rotational analyses on spectroscopic experiments. The analytical function is of much realistic importance since it is possible to be applied to predict fine transitional structure and study reaction dynamic process.
\end{abstract}
\end{frontmatter}

\section{Introduction}
The transition metal monohydrides and monohalides have been extensively studied over several decades because they are of considerable interest in various fields such as astrophysics, catalytic chemistry, high-temperature chemistry and surface material\cite{cc,as,bl}. In particular diatomic molecules of La are of great importance as test cases for modeling the role that \textit{d} electrons play in chemical bonds, since lanthanum has the simplest open \emph{d} shell electronic configuration [(core)$5d^16s^2$]. There have been significant efforts made in both spectroscopic investigations and \emph{ab initio} calculations in order to get a better understanding of the electronic structure of such molecules. The visible band systems of monohalides LaH, LaF, LaCl are observed and rotationally analyzed\cite{basi,barr,bedv,kkh,jxin}. A number of theoretical calculations also exist using ligand field theory and \emph{ab initio} methods at all levels\cite{kmh,kds,hdl,ljvs}.

However up to now theoretical studies mainly focus on molecular equilibrium structure. Only several papers were devoted to calculations of global potential energy curves of LaH and LaF\cite{kb,chsh,csxs}, which is not compatible with so many experimental spectroscopic data. In the case of LaCl, there is even no report on theoretical calculation of its structure and molecular properties. Global potential energy curve is significant for the investigation of collision reaction dynamics when these compounds are used as catalyst. In the present paper potential energy curve and analytical potential energy function, as well as some molecular properties for LaCl molecule are firstly calculated and presented with energy-consistent Relativistic Effective Core Potentials (RECPs) and the corresponding valence basis sets including diffuse and polarization functions.

\section{Calculation Method}
Quantum chemical calculations of lanthanide complexes are relatively difficult due to complex electronic structure and large systems. A large number of low-lying electronic states of different spatial and spin symmetries may mix heavily. Electron correlation and relativistic effects can be very considerable, and spin-orbit interactions are often quite important for these systems. Accurate theoretical calculations have to take account of all these effects properly. As for the ground state $X^1\Sigma^+$ of LaCl, first-order spin-orbit interactions needn't to be considered. Although in recent years some advance has been achieved in four-component all-electron relativistic calculations\cite{kds,hdl,ljvs}, they are too costly. We adopt relativistic effective core potential approximation, developed in 1970s and applied successfully to many heavy-metal compounds (e.g.\ \cite{kds}), to reduce the enormous calculation amount. Energy-consistent RECPs delivered by Stuttgart/Dresden group\cite{dssp} are utilized, where 46 core electrons of La are replaced by effective potentials derived from quasi-relativistic atomic wave functions. Only the outer 11 electrons $5s^25p^65d^16s^2$ need to be treated explicitly in the calculation. Valence basis sets are contracted as (8s7p6d4f2g)/[6s5p4d4f2g] including diffuse functions 1s1p1d and polarization functions 4f2g\cite{dssp}. The exponents of the diffuse functions are determined according to even-tempered principle, and 4f2g's exponents are 4.5, 1.5, 0.5, 0.167, 1.5 and 0.5 respectively. For Cl atom Dunning's correlation consistent triple-zeta basis sets augmented with diffuse functions, i.e.\ aug-cc-pVTZ, are adopted. We use hybrid density functional B3LYP, Moller-Plesset correlation energy correction MP2 and quadratic configuration interaction method QCISD(T) to account for electron correlation effects. Basis Set Superposition Error (BSSE) is corrected by Boys and Bernardi's Counter-Poise procedure\cite{bb}. All the calculations are completed on SGI 1450 server with 4GB memory by Gaussian 98 package\cite{g98}.

\section{Equilibrium Geometry}
The possible electronic state for the ground state of LaCl may be the singlet $^{1}\Sigma^{+}$ or the triplet $^{3}\Delta$. These two states are both optimized using B3LYP of Density-Functional Theory (DFT). The bond length is 0.2520 nm and 0.2576 nm for the singlet and triplet respectively, and the corresponding experimental values obtained from spectroscopic constant $B_{e}$ are 0.2498 nm and 0.2547 nm. With optimized geometries QCISD(T) is used to perform high accuracy energy calculation. The calculated energy of singlet is about 0.15 eV lower than that of triplet, so the ground electronic state is determined to be the singlet. This is in accordance with the assignment of \cite{jxin} although there was no compellent experimental evidence in that paper. Theoretical calculations here predict LaCl has analogic ground electronic configuration with LaH and LaF, which is still to be confirmed by future experiments. 

Table \ref{geodis} are bond length, bond dissociation energy and harmonic frequency obtained with DFT and post-HF methods. Compared with experimental values, the bond length is 2.2 pm, 1.6 pm larger at B3LYP, QCISD(T) levels and 0.6 pm smaller at MP2 level. The bias for harmonic frequency is about 10 cm$^{-1}$, 1 cm$^{-1}$ and 12 cm$^{-1}$ respectively. With respect to bond dissociation energy, the values of B3LYP and QCISD(T) nearly equal (4.982 eV and 5.045 eV), but MP2 value is about 1.8 eV larger. Though by far there is no experimental report on the dissociation energy, our calculations estimate it is near 5.0 eV. This estimation is a little smaller than 5.41 eV, predicted by Kaledin et al.\ \cite{khf} with a statistic argument that thermochemical properties of the lanthanide monohalides may have simple linear relationships. Since B3LYP and QCISD(T) can provide more perfect results of energy, we employ them to calculate the potential energy curve for the ground state of LaCl.

\section{Potential Energy Curve and Analytical Function}
In order to calculate potential energy curve, first all the possible electronic states and dissociation limits of LaCl should be determined based on atomic and molecular reaction statics\cite{zhu,ms}. Firstly the separated atomic group me\-thod can be used to derive the possible electronic state for the ground state. The ground electronic states of La and Cl are $^{2}D_g$ and $^{2}P_u$ respectively, both belong to SU(n) group. The symmetry is degraded to $C_{\infty{V}}$ when La and Cl approach and form the molecule LaCl. So representations $^{2}D_g$ and $^{2}P_u$ of group SU(n) can be resolved into direct sum of the representations of group $C_{\infty{V}}$ as $$^2D_g=\/^2\Sigma_g^{+}\oplus{^2\Pi_g}\oplus{^2\Delta_g},\hspace{3mm}^2P_u=\/^2\Sigma_u^{+}\oplus{^2\Pi_u}.$$
The direct product of atomic representations $^{2}D_g$ and $^{2}P_u$ is reduced as $$(\/^2\Sigma_g^{+}\oplus{^2\Pi_g}\oplus{^2\Delta_g})\otimes(\/^2\Sigma_u^{+}\oplus{^2\Pi_u})=\/^{1,3}\Sigma^{+}(2)\oplus{^{1,3}\Sigma^{-}}\oplus{^{1,3}\Pi(3)}\oplus{^{1,3}\Delta(2)}\oplus{^{1,3}\Phi}.$$
Obviously the possible electronic states for LaCl are $^{1,3}\Sigma^{+}$, $^{1,3}\Sigma^{-}$, $^{1,3}\Pi$, $^{1,3}\Delta$, $^{1,3}\Phi$. Secondly the calculated electronic configuration of LaCl is $\sigma^2\sigma^2\sigma^2\pi^4\sigma^2\sigma^2$ $\pi^4\sigma^2\sigma^2\pi^4\sigma^2$, which belongs to the state $^{1}\Sigma^{+}$. Therefore the ground state of LaCl is $X^{1}\Sigma^{+}$. According to the principle of microscopic reversibility, the dissociation limit for the ground state $X^{1}\Sigma^{+}$ can be expressed as
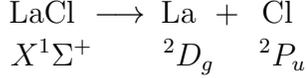
\begin{eqnarray*}
\mathrm{LaCl}\ \longrightarrow &{\ } \mathrm{La}\ \;+\: &{\ } \mathrm{Cl} \\ 
X^{1}\Sigma^{+}\hspace{7.0mm} & ^{2}D_g\hspace{2.6mm} & \hspace{1.0mm}^{2}P_u
\end{eqnarray*}

\begin{figure}
\begin{center}
\includegraphics[scale=0.6,angle=-90]{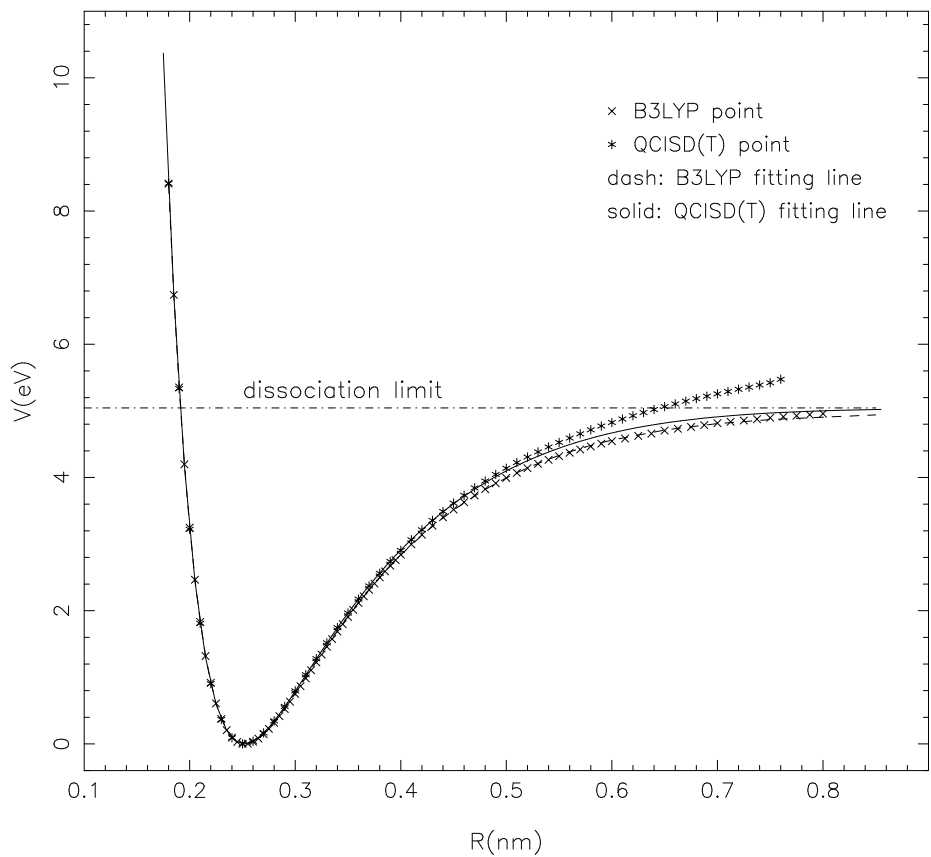}
\caption{\label{fig}Potential energy curves for the ground state $X^{1}\Sigma^{+}$ of LaCl. The dash and solid lines are respectively fitting lines of B3LYP and QCISD(T).} \end{center}
\end{figure}

Potential energy scans are carried out every 5 pm for a calculation point while the internuclear distance is between 0.18 nm and 0.8 nm. Potential energy curves we got are shown in Fig \ref{fig}. Murrell-Sorbie (M-S) function\cite{ms} is one of the best analytical functions which can describe both short-range and long-range interaction very well. Its form reads
$$V(R)=-D(1+a_1\xi+a_2\xi^2+a_3\xi^3)\mathrm{exp}(-a_1\xi)+D,$$
where $\xi=R/R_e-1$, $R$ and $R_e$ are internuclear distance and its equilibrium value, $D$ is bond dissociation energy. In addition M-S function can be expanded to the following approximate power series at equilibrium position as what Dunham did in 1932\cite{dunham}, i.e.\
$$V(R)=a_0\xi^2(1+a_1\xi+a_2\xi^2+a_3\xi^3+a_4\xi^4+a_5\xi^5+a_6\xi^6).$$
The coefficients $a_1,\cdots,a_6$ shown in table \ref{potfun}, are calculated by a nonlinear least-square fitting of the data of potential energy curve. The formulae to obtain force constants from M-S analytical function are $f_2=D(a_1^2-2a_2)$, $f_3=D(-2a_1^3+6a_1a_2-6a_3)$ and $f_4=D(3a_1^4-12a_1^2a_2+24a_1a_3)$. Standard Dunham treatment\cite{dunham} is adopted to calculate vibrotational spectroscopic constants, which are listed in table \ref{vibcon} accompanied with those of rotational analyses from experimental spectra.

BSSE is about 0.12 eV for QCISD(T) method with large basis sets in this paper. Considering BSSE correction is often necessary in the high accuracy calculation of global potential energy curve. In Fig \ref{fig} potential energy curves obtained by B3LYP and QCISD(T) are almost the same in the range from 0.18 nm to 0.4 nm. From 0.4 nm two curves begin to separate. The difference is about 0.5 eV when the internuclear distance is 0.8 nm. Comparing with the value of bond dissociation energy in table \ref{geodis}, we can find potential energy curve got by QCISD(T) does not meet the dissociation limit of ground state, although BSSE correction has been made. This problem has been referred to by Wang et al.\cite{wgsw}. They argued that the denominator of perturbation energy term of single-reference ab initio methods such as QCISD(T) and CCSD(T) is connected with the energy gap between Highest Occupied Molecuar Orbit (HOMO) and Lowest Unoccupied Molecular Orbit (LUMO). It was found that HOMO-LUMO gap becomes small when the bond separation is far from equilibrium, so that the perturbation energy is somewhat large and finally results in the too high potential energy. Another possible reason is that RECPs and the corresponding valence basis sets are optimized at equilibrium geometry. Multi-reference configuration interaction or multi-reference perturbation theory is expected to be efficiently applied to such system far from $R_e$.

All the parameters for M-S analytical function and Dunham expansion are given in table \ref{potfun}. The fitted dissociation energy is 4.956 eV for B3LYP method, which is very close to the value 4.982 eV calculated at equilibrium position. Since potential energy of large distance is 0.5 eV above dissociation limit, we choose the data of QCISD(T) of $R<0.4$ nm to fit the M-S analytical function. Thus the fitted dissociation energy 5.105 eV becomes very close to 5.045 eV in table \ref{geodis}. Using the analytical function, most of the calculated vibrotational constants except $\omega_ey_e$, $\beta_e$ are in rather good agreement with those measured in thermal emission spectroscopy experiments\cite{jxin}. For example with B3LYP method, the vibrational frequency $\omega_e$ for the ground state $X^{1}\Sigma^{+}$ is 338.7 cm$^{-1}$, and experimental value is 341.6 cm$^{-1}$. The discrepancy with harmonic frequency 332.0 cm$^{-1}$ of table \ref{geodis} is because of disharmonic effect of electron vibrations. Two other spectroscopic constants $\omega_ez_e$ and $H_e$ that the experiment didn't analyze out can be also given on theory, the values of B3LYP coincide well with those of QCISD(T). With these spectroscopic constants, the vibrotational energy levels for the ground state of molecule LaCl may be expressed as $[\omega_e-\omega_e\chi_e(\nu+0.5)](\nu+0.5)+[B_e-\alpha_e(\nu+0.5)]J(J+1)-[D_e+\beta_e(\nu+0.5)]J^{2}(J+1)^{2}+H_eJ^{3}(J+1)^{3}$, where $\nu$ and $J$ correspond to vibration and rotation quantum number respectively.

\section{Summary and Discussion}
The electronic state and dissociation limit of the ground state of lanthanum monochloride have been deduced according to the principles of atomic and molecular reaction statics and quantum chemical calculations. Molecular properties at equilibrium position have also been firstly calculated on B3LYP, MP2 and QCISD(T) levels with relativistic effective core potential approximation. Our theoretical calculations show LaCl has analogic ground state $X^1\Sigma^+$ with LaH and LaF, which is to be confirmed by future experiments. Potential energy curves and their M-S analytical functions and Dunham expansions have been obtained with B3LYP and QCISD(T) methods. The calculated vibrotational spectroscopic constants from analytical functions are in very good agreement with experiment results.

All the quantum chemical calculations demonstrate DFT and QCISD(T) under energy-consistent RECPs approximation are efficient and reliable to evaluate relativistic and electron correlation effects. If consider only bond dissociation energy, the results of B3LYP and QCISD(T) are totally better than those of MP2. We estimate the dissociation energy of gound state is about 5.0 eV, smaller than the prediction of \cite{khf}. Accurate experimental value is expected. The potential energy curve of QCISD(T) at large diatomic distance is 0.5 eV above dissociation energy, so it doesn't satisfy the requirement of dissociation limit. We propose the scheme that fit the data of DFT or the data around equilibrium position of QCISD(T) to derive M-S analytical function for potential energy curve. According to this scheme, the calculated analytical functions of B3LYP and QCISD(T) are both consistent with the dissociation limit of the ground state of LaCl molecule.

The agreement of spectroscopic constants of theoretical calculations and experimental measurements shows that Murrell-Sorbie potential function we got is a good analytical function to express the potential energy curve of lanthanum monochloride. The analytical function can predict transitional frequency and the intensity in the fine structure of experimental spectra. Its application to the investigation of detail collision reaction dynamic processes is expected when LaCl is used as catalyst. Thus the analytical function is of much realistic importance. The calculation scheme that obtain potential energy curve under RECPs approximation and how to use Murrell-Sorbie analytical function to fit the potential energy curve can be extended to other transition heavy-metal compounds.

\section*{Acknowledgements}
L.H. Chen thanks Prof. Michael Dolg at Bonn University for supporting energy-consistent RECPs of La atom and Prof. Hermann Stoll at Stuttgart University for helpful suggestions.

\clearpage

\begin{table}
\caption{\label{geodis}Geometry, bond dissociation energy and harmonic frequency of LaCl}
\scriptsize
\begin{tabular}{cccccccc} \hline
Method & $R_e$(nm) & $D$(eV) & $\nu$(cm$^{-1}$) & Method & $R_e$(nm) & $D$(eV) & $\nu$(cm$^{-1}$) \\ \hline
B3LYP & 0.2520 & 4.982 & 332.0 & QCISD(T) & 0.2514 & 5.045 & 342.7 \\
MP2 & 0.2492 & 6.788 & 353.3 & Exp. & 0.2498$^{\rm a}$ & - & 341.6$^{\rm b}$ \\ \hline
\end{tabular}
\\ $^{\rm a}$ Converted from experimental $B_e$ of \cite{jxin}.
\\ $^{\rm b}$ Experimental vibration frequency taken from \cite{jxin}.
\normalsize
\end{table}

\begin{table}
\caption{\label{potfun}Parameters and force constants$^{\rm a}$ of analytical potential function for the ground state $X^{1}\Sigma^{+}$ of LaCl}
\footnotesize
\begin{tabular}{cccccccc} \hline
M-S & $D$(eV) & $a_1$ & $a_2$ & $a_3$ & $f_2$ & $f_3$ & $f_4$\\ \hline
B3LYP & 4.956 & 4.019 & 0.5252 & 4.916 & 0.189 & -7.279 & 227.51 \\ 
QCISD(T) & 5.105 & 4.025 & 0.6610 & 4.980 & 0.192 & -7.426 & 233.25\\ \hline
Dunham & $a_0$(eV) & $a_1$ & $a_2$ & $a_3$ & $a_4$ & $a_5$ & $a_6$ \\ \hline
B3LYP & 37.958 & -3.218 & 6.194 & -8.573 & 9.151 & -7.870 & 5.630 \\
QCISD(T) & 37.970 & -3.233 & 6.384 & -9.189 & 10.271 & -9.275 & 6.974 \\ \hline
\end{tabular}
\\ $^{\rm a}$ Units for force constants $f_2$, $f_3$, $f_4$ are fJ$\cdot$nm$^{-2}$, fJ$\cdot$nm$^{-3}$, fJ$\cdot$nm$^{-4}$ respectively.
\normalsize
\end{table}

\begin{table}
\caption{\label{vibcon}Vibrotational constants$^{\rm a}$ for the ground state $X^{1}\Sigma^{+}$ of LaCl}
\scriptsize
\begin{tabular}{cccccccccc} \hline
Item & $\omega_e$ & $10^{2}B_e$ & $\omega_e\chi_e$ & $10^4\omega_ey_e$ & $10^{4}\alpha_e$ & $10^{11}\beta_e$ & $10^{8}D_e$ & $10^6\omega_ez_e$ & $10^{15}H_e$ \\ \hline
B3LYP & 338.7 & 9.503 & 0.9588 & 7.88 & 3.579 & 5.11 & 2.991 & 2.72 & -2.00 \\
QCISD(T) & 342.0 & 9.546 & 0.9571 & 8.35 & 3.571 & 4.76 & 2.975 & 2.52 & -2.16 \\
Exp.$^{\rm b}$ & 341.6 & 9.670 & 0.9797 & 7.46 & 3.640 & 6.91 & 3.119 & - & - \\ \hline
\end{tabular}
\\ $^{\rm a}$ Unit is cm$^{-1}$.
\\ $^{\rm b}$ Taken from \cite{jxin}.
\normalsize
\end{table}
\clearpage

\end{document}